\documentstyle[12pt,epsf]{article}

\newcommand{\la}[1]{\label{#1}}

\newcommand{\be}{\begin{equation}}
\newcommand{\ee}{\end{equation}}
\newcommand{\ba}{\begin{eqnarray}}
\newcommand{\ea}{\end{eqnarray}}
\newcommand{\bi}{\begin{itemize}}
\newcommand{\ei}{\end{itemize}}
\newcommand{\rmi}[1]{{\mbox{\scriptsize #1}}}

\newcommand{\nr}[1]{(\ref{#1})}
\newcommand{\bfx}{\mbox{\bf x}}
\newcommand{\x}{{\bf x}}

\newcommand{\fr}[2]{{\frac{#1}{#2}}}
\newcommand{\msbar}{\overline{\mbox{\rm MS}}}

\def\lsi{\raise0.3ex
\hbox{$<$\kern-0.75em\raise-1.1ex\hbox{$\sim$}}}
\def\gsi{\raise0.3ex
\hbox{$>$\kern-0.75em\raise-1.1ex\hbox{$\sim$}}}

\newcommand{\gsim}{\mathop{\gsi}}

\begin{document}

\sloppy
\begin{titlepage}

\begin{flushright}
CERN-TH/98-70\\
HIP-1998-14/TH\\
hep-ph/9803367\\
March 1998
\end{flushright}
\begin{centering}
\vfill
{\bf THERMODYNAMICS OF GAUGE-INVARIANT U(1) VORTICES \\
FROM LATTICE MONTE CARLO SIMULATIONS}

\vspace{0.4cm}
K. Kajantie$^{\rm a,b,}$\footnote{keijo.kajantie@cern.ch},
M. Karjalainen$^{\rm b,}$\footnote{mika.karjalainen@helsinki.fi},
M. Laine$^{\rm a,b,}$\footnote{mikko.laine@cern.ch},
J. Peisa$^{\rm c,}$\footnote{peisa@amtp.liv.ac.uk}, and
A. Rajantie$^{\rm d,}$\footnote{arttu.rajantie@helsinki.fi} \\

\vspace{0.4cm}
{\em $^{\rm a}$Theory Division, CERN, CH-1211 Geneva 23, Switzerland}

\vspace{0.1cm}
{\em $^{\rm b}$Department of Physics,
P.O.Box 9, 00014 University of Helsinki, Finland}

\vspace{0.1cm}
{\em $^{\rm c}$Department of Mathematical Sciences,
University of Liverpool, \\
Liverpool L69 3BX, UK}

\vspace{0.1cm}
{\em $^{\rm d}$Helsinki Institute of Physics,
P.O.Box 9, 00014 University of Helsinki, Finland}

\vspace{5mm}
{\bf Abstract}

\end{centering}

\vspace{0.3cm}\noindent
We study non-perturbatively and from first principles
the thermodynamics of vortices in 3d U(1) gauge+Higgs theory, 
or the Ginzburg-Landau model, which has frequently been used
as a model for cosmological topological defect formation. 
We discretize the system and introduce a gauge-invariant definition 
of a vortex passing through a loop on the lattice. We then study
with Monte Carlo simulations the total vortex density, extract
the physically meaningful part thereof, and demonstrate that it
has a well-defined continuum limit. The total vortex density 
behaves as a pseudo order parameter, having a discontinuity 
in the regime of first order transitions and behaving
continuously in the regime of second order transitions. Finally, 
we discuss further gauge-invariant observables to be measured.

\vspace{3mm}\noindent

\vfill \vfill
\end{titlepage}

\section{Introduction}

Vortices play a significant role
from the low temperatures of liquid crystals~\cite{lcexperiments},  
superfluids~\cite{rephe}
and high-$T_c$ superconductors~\cite{reptc}, to the 
relativistic temperatures of the Early Universe~\cite{repco}.
In low temperature systems, vortices can be
directly observed~\cite{lcexperiments,scexperiments}; 
in cosmology, one has studied their effects
on the inhomogeneities leading to 
structure formation~\cite{kibble}.
Consequently, vortices have been a subject of immense 
interest during the last few years. Nevertheless, 
some important questions, related in particular
to non-perturbative studies of vortices in gauge 
theories, remain poorly understood. 

Among the most fundamental principles of Nature appear
to be gauge invariance and spontaneous
symmetry breaking, and already 
the simplest theory with these properties, 
the locally U(1) symmetric gauge+scalar 
quantum field theory or the Ginzburg-Landau (GL) model, 
contains vortices.
The GL model does describe real physics
in liquid crystals and superconductors~\cite{kleinert}, while in cosmology
it is to be viewed as a simple toy model. The phase structure
of the GL model is non-trivial: in the type I regime, 
there is a first order transition \cite{hlm}, whereas in the type II
regime, the transition is assumed 
to be of the second order \cite{mr}--\cite{ht}. 

One of the mentioned open questions 
arises immediately when one realizes that the type II regime
of the GL model is completely non-perturbative: perturbation 
theory does not describe the transition at all~\cite{dh}. 
The only known systematic and controllable method for studying 
this regime are lattice Monte Carlo 
simulations~\cite{dh}--\cite{u1big}. 
Yet as to date, to our knowledge, 
the vortex density has not been studied in detail
on the lattice even in thermodynamical equilibrium.
The purpose of this paper is (a) to provide a
gauge-invariant formulation for studying vortices on the lattice,
(b) to measure the vortex density both in type I and type II
regimes, and (c) to extrapolate the results to the infinite volume and
continuum limits. The length distribution of vortices
will be studied in a future publication~\cite{a6}.
In a U(1) scalar field theory without 
gauge symmetry, the thermodynamics of vortices has 
previously been addressed in~\cite{abh}.

Let us stress that considering the thermodynamics 
of vortices is certainly only a starting point. 
Ultimately one is interested in the real-time scaling 
properties of vortex networks created in a non-equilibrium 
situation~(see, e.g., \cite{tvreview}--\cite{vah}).
However, the thermodynamical equilibrium situation 
provides the initial conditions for such non-equilibrium
processes. Furthermore, it is clear that non-equilibrium 
physics cannot be understood in quantitative detail before 
the equilibrium limit is under control.

\section{The theory in the continuum and on the lattice}

Let us start by defining the theory.
The continuum theory is defined by the functional integral
\ba
Z&=&\int {\cal D}A_i{\cal D}\phi\,\exp\bigl[-S(A_i,\phi)\bigr], \la{z} \\
S&=&\int d^3x\biggl[\fr14 F_{ij}^2+
|D_i\phi|^2 
+m_3^2 \phi^*\phi + \lambda_3 \left(\phi^*\phi\right)^2\bigg], 
\label{action}
\end{eqnarray}
where $F_{ij}=\partial_iA_j-\partial_jA_i$ 
and $D_i=\partial_i+ie_3A_i$.
The theory is invariant under the gauge transformations 
\be
\phi(\x)\to e^{i\theta(\x)}\phi(\x),\quad
A_i(\x)\to A_i(\x)-\partial_i\theta(\x)/e_3.
\la{gt1}
\ee
Writing $\phi(\bfx)=v(\bfx)\exp[i\gamma(\bfx)]$, the first
of these can be rewritten as 
$\gamma(\bfx)\to [\gamma(\bfx)+\theta(\bfx)]_\pi$,
where $[X]_\pi\equiv X+2\pi n$ such that $[X]_\pi\in (-\pi,\pi]$
and we have chosen to represent the phase of $\phi$
by a number in this interval.
The theory in eq.~\nr{action}
is parameterized by the scale $e_3^2$ and by the 
two dimensionless
ratios
\be
y= { m_3^2(e_3^2)\over e_3^4},\quad x={\lambda_3\over e_3^2},
\la{parameters}
\ee
where $m_3^2(\mu)$ is the mass parameter in 
the $\msbar$ dimensional regularization scheme in 
$3-2\epsilon$ dimensions. Expressions for $x$ and $y$ in 
terms of the original physical parameters of both  
4d high temperature scalar+fermion electrodynamics
and 3d low-$T_c$ superconductivity, 
have been discussed in~\cite{u1big} where we refer to
for more details. Here we just study the 
theory as a function of $x,y$. The phase diagram
is shown in Fig.~\ref{ycx}.

\begin{figure}[t]
\hspace{1cm}
\epsfxsize=8cm
\centerline{\epsffile{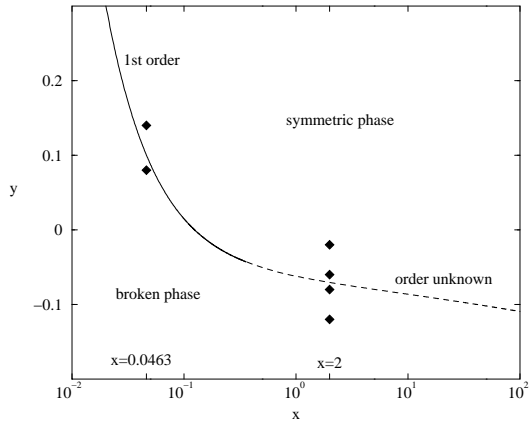}}

\vspace*{-0.5cm}

\caption[a]{The phase diagram of the GL model,
together with the main simulation points
(marked with the diamonds).}
\la{ycx}
\end{figure}

As is well known, the classical counterpart of the
theory in eq.~\nr{action} 
admits vortex, or string solutions, in the broken 
(superconducting) phase. 
In other words, the classical
equations of motion have solutions in which the symmetry 
is restored at the core, but is broken far away from the core~\cite{no}. 
The existence of a vortex inside a loop $C$ can be identified
by computing the line integral 
\be
n_C = \frac{1}{2\pi}\oint_C \! d\bfx\cdot\nabla \gamma(\bfx), 
\la{nC}
\ee
where the winding number 
$n_C$ is an integer,
and a non-zero $n_C$ signals a vortex inside the loop. 
It can be seen from eq.~\nr{gt1} that $n_C$
is gauge-invariant for single-valued gauge transformations $\theta(\bfx)$. 
Thus vortices are
physical objects, which can be generated in a phase transition
or by applying non-trivial boundary conditions,
and they can be observed in the superconducting phase. 
However, in the equilibrium state they are also created and destroyed
by thermal fluctuations, and $\langle|n_C|\rangle$ 
gives their density both in the
broken and in the symmetric phase. Moreover, it is expected that the
behavior of vortices is qualitatively different in the two phases.
As the dynamics is non-perturbative, 
it is then clear that it is very important to be able to 
study vortices with lattice simulations.  

It should be noted here that one often assumes that in the regime
of large $x$, the gauge fields are not essential and
one can approximate the theory in eq.~\nr{action}
by the 3d XY-model with a global U(1) symmetry, or its dual version
(see, e.g., \cite{dh,kovneretal,kksfort})
whose fundamental objects are the vortices.
While these approximations simplify the problem significantly,
their validity is uncertain for finite $x$.
Thus we consider it essential to approach the 
problem directly with the original theory in eq.~\nr{action}.

To allow for lattice simulations, the theory in eq.~\nr{action}
has to be discretized. As usual, we 
introduce the link field $U_i(\bfx)=
\exp[iae_3A_i(\bfx)]\equiv\exp[i\alpha_i(\bfx)]$. 
Rescaling the continuum scalar field to a 
dimensionless lattice field by 
$\phi^*\phi\to \beta_H\phi^*\phi/2a$,
the lattice action becomes 
\ba
S & = & \beta_G \!\!\sum_{\bfx,i<j}
\fr12\hat{F}_{ij}^2(\bfx)
  -\beta_H \sum_{\bfx, i} {\mbox{Re}}\, 
\phi^*(\bfx) U_i(\bfx)\phi(\bfx+\hat{i})\nonumber\\   
  & & +\sum_{\bfx} \phi^*(\bfx) \phi(\bfx)
+\beta_R\sum_{\bfx} \left[\phi^*(\bfx)\phi(\bfx)-1\right]^2, 
 \la{standardaction} 
\ea
where 
$\hat{F}_{ij}(\bfx)=\alpha_i(\bfx)+\alpha_j(\bfx+\hat i)-
\alpha_i(\bfx+\hat j)-\alpha_j(\bfx)$. 
We use here the non-compact formulation
for the gauge fields. The gauge transformation properties 
in eq.~\nr{gt1} go over into
\ba
\gamma(\bfx)  &\to&  [\gamma(\bfx)+\theta(\bfx)]_\pi,\nonumber \\
\alpha_i(\bfx) &\to& 
\alpha_i(\bfx)+\theta(\bfx)-\theta(\bfx+\hat{i}). \la{gt3}
\ea
Discretization can be viewed as a different
regularization scheme, and in order to describe the same 
continuum physics as in eq.~\nr{action}, one has to make 
a 2-loop computation relating the counterterms~\cite{lattcont}. 
As a result, the lattice couplings $\beta_G, \beta_H, \beta_R$ are 
determined from 
\ba
& & \beta_G={1\over e_3^2a},\qquad \beta_R={x\beta_H^2\over4\beta_G},
\la{betaR} \\
& & 2\beta_G^2\left({1\over\beta_H}-3-{x\beta_H\over2\beta_G}\right)
=y-{3.1759115(1+2x)\beta_G\over2\pi}\la{betah}\\
&&\hspace*{1cm} -\frac{
(-4+8x-8x^2)(\log6\beta_G+0.09)-1.1+4.6x}{16\pi^2}.\nonumber
\ea
Thus for a given continuum theory depending on one scale $e_3^2$
and the two dimensionless 
parameters $y,x$,  the use of a lattice introduces a regulator
scale $a$, and eqs.~\nr{standardaction}--\nr{betah}  
specify, up to terms of order $e_3^2a$, 
the corresponding lattice action. Note that the simulations
in~\cite{ncdef} correspond to $\beta_R\to\infty$, 
whereas according to eq.~\nr{betaR}, $\beta_R\to 0$ in 
the continuum limit for any finite $x$ ($\beta_H\to1/3$).

\section{Gauge-invariant vortices}

Consider now vortices. The naive discretization 
of eq.~\nr{nC} gives  the standard algorithm used 
in scalar theories without gauge fields. 
For each loop $C$ one would define the winding number $\tilde n_C$ 
of the phase $\gamma$
of the scalar field.
However, any $\gamma(\bfx)$ can be changed arbitrarily 
with a gauge transformation, see eq.~\nr{gt3}. Thus 
the $\gamma(\bfx)$'s are essentially random numbers, 
and $\tilde n_C$
does not contain any real 
information about the dynamics of the system.
Indeed, the result for a loop around a single plaquette,
$\tilde N_{1\times1} \equiv \langle|\tilde n_{1\times 1}(\x)|\rangle$,
would equal $1/3$ in the case of completely uncorrelated field values, 
and this is 
what we measure from lattice simulations for $\tilde N_{1\times1}$, 
irrespective of the parameters of the theory (to be more
precise, we always get $\tilde N_{1\times1} = 0.32(1)\ldots0.33(1)$). 
Thus the quantity $\tilde n_C$ has to be rejected.

One solution sometimes used in the literature
would be to fix the gauge, which makes $\tilde n_C$ non-trivial.
However, its value depends crucially on the gauge chosen, and it is even 
possible to choose a gauge in which $\tilde n_C$ always vanishes.
Therefore, we believe that it is important to use
an explicitly gauge-invariant definition, in order
to be able to interpret the results correctly. 

Fortunately, the problem with $\tilde n_C$
is not one of principle, and 
a satisfactory definition can be given. 
For each positively 
directed link
$l=(\x,\x+\hat i)$ let us define 
\be
Y_{(\x,\x+\hat i)}=[\alpha_i(\x)+\gamma(\x+\hat i)-
\gamma(\x)]_\pi-\alpha_i(\x).
\ee
For links with negative direction the sign of $Y_l$ is changed:
$
Y_{(\x,\x-\hat i)}=-Y_{(\x-\hat i,\x)}.
$
Then, for each closed loop $C$, we can define
\be
\la{correct}
Y_C=\sum_{l\in C}Y_l\equiv 2\pi n_C.
\ee
This definition has four main properties:

\noindent
(a) For any field configuration and any loop $C$, $n_C\in {\sf Z\!\!Z}$.

\noindent
(b) Directly from eqs.~\nr{gt3}, one can see that the part of $Y_l$
in the square brackets is gauge-invariant. The term $-\alpha_i(\x)$ is not,
but when summed over a closed loop 
into $Y_C$, the gauge dependence cancels. Hence
$Y_C$ is gauge-invariant.

\noindent
(c) Since $Y_C$ is gauge-invariant, one can always tune the 
gauge used in the evaluation of the $Y_l$'s such that the fields
appearing are perturbatively small. But then, in the continuum limit, 
$\alpha_i = e_3 a A_i$ goes to zero and 
one gets the correct continuum limit containing
only the phase of $\phi$.

\noindent
(d) The quantity $Y_C$ is additive: if there is a loop $C$
consisting of the loops $A,B$, then $Y_C = Y_A+Y_B$. This is what one
would require for counting the number of vortices going through loops
of different sizes. This also implies that vortex lines cannot end, and 
therefore they form closed vortex loops. Note that we use a non-compact
gauge field so that there are no monopoles.

Based on these properties, eq.~\nr{correct}
provides a valid formulation for counting vortices in the 
locally symmetric U(1) theory~\cite{artun}.
This definition of the winding number coincides with that given in 
Ref.~\cite{ncdef} for the case $\beta_R=\infty$.

\section{Simulations and results}

In order to see how the gauge-invariant definition 
performs in practice, we have made lattice Monte Carlo 
simulations in the GL model. We choose $C=n\times n\equiv$
a loop around a plaquette of size $n\times n$
in eq.~\nr{correct}, and measure
\be
N_{n\times n} \equiv \langle |n_{n\times n}|\rangle.
\ee
The quantity $N_{n\times n}$ measures the average net number of
vortices through a loop of size $n\times n$, irrespective
of the net direction. Keeping track of the 
direction would give zero: for symmetry reasons, 
$\langle n_{n\times n}\rangle=0$. 
In practice, we average $\langle |n_{n\times n}|\rangle$
over all lattice sites and directions, to improve on 
the statistics.

Simulations
are made at two values of $x$: $x=0.0463$ corresponds to
a strongly type I superconductor, $x=2$ to a strongly type II
superconductor. For each $x$, values of $y$ are chosen on 
both sides of the transition (see Fig.~\ref{ycx}). For each
such continuum parameter point, several lattice spacings 
are chosen: $\beta_G=1/(e_3^2 a)=1,2,3,4,6,8,12$. For $\beta_G=4$
($x=0.0463$) and $\beta_G=1,2$ ($x=2$), several volumes are 
chosen, in order to test that the finite volume 
effects are small.
The volumes thus arrived at (24$^2\times$48 for $\beta_G=4$
both at $x=0.0463,x=2$) are then 
scaled with $\beta_G$ such that the physical volume 
(in units of $1/e_3^6$)
remains constant. The orders of magnitude 
for $\beta_G$ and the volume come from 
the requirement that the physical correlation lengths, 
of order $(0.5\ldots2)/e_3^2$ at $x=0.0463$ and 
$(1.5\ldots 3)/e_3^2$ at $x=2$~\cite{prb}, are much longer than the 
lattice spacing but much shorter than the extent of 
the whole lattice (with the exception of the photon 
in the symmetric phase).

\begin{figure}[h]
\hspace{1cm}
\epsfysize=8cm
\centerline{\epsffile{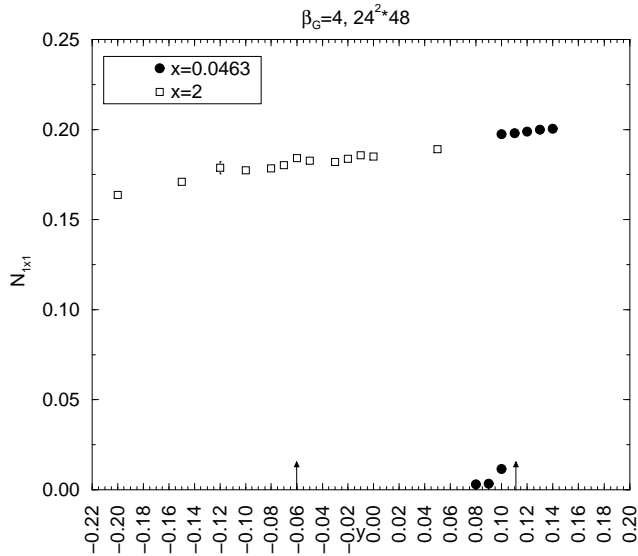}}
 
\caption[a]{The values of $N_{1\times 1}$
at $x=0.0463$ and $x=2$, as a function of $y$. The transition
points are indicated by the arrows. There is a discontinuity 
at the first order transition at $x=0.0463$, whereas in the 
regime of large $x$,  $N_{1\times 1}$ behaves continuously.
Some of the points at $x=0.0463$ correspond to metastable phases.}  
\la{yij}
\end{figure}

The results for $N_{1\times 1}$
as a function of $y$ are shown in Fig.~\ref{yij}. 
The two values of $x$ have a different critical point $y_c$, 
see Fig.~\ref{ycx}. At small $x$
the vortex density is very small in the broken phase but jumps 
discontinuously to a large value at the transition. However, the
symmetric-phase value $\approx 0.2$ is much smaller than the trivial
value $1/3$. At large $x$ the behavior is completely different.
The total vortex density is large also rather deep in the broken phase and
there is no discontinuity at the transition. 

\begin{figure}[t]
\hspace{1cm}
\epsfysize=8cm
\centerline{\epsffile{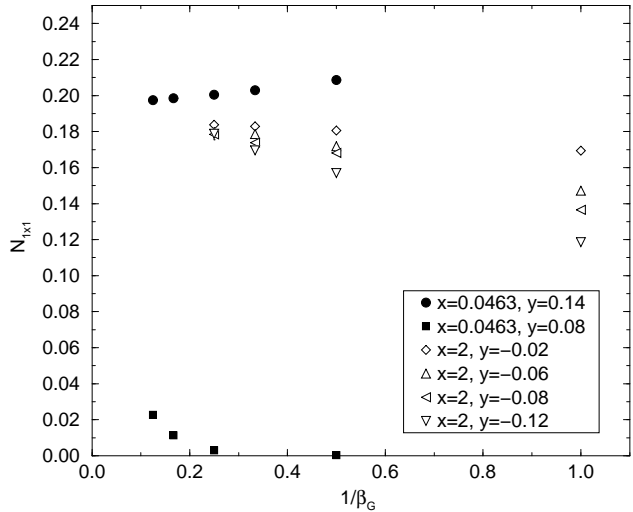}}
\caption[a]{The continuum extrapolation of $N_{1\times 1}$
           for different values of $x,y$. Even though 
           there is a lot of structure at finite $\beta_G$, 
           all results approach a single point in the 
           continuum limit $\beta_G\to\infty$, 
           including those in the broken phase.
           Both polynomial ($c_0+c_1/\beta_G+c_2/\beta_G^2$)
           and logarithmic ($c_0+c_1\ln\beta_G/\beta_G+c_2/\beta_G$)
           fits are allowed at $1/\beta_G\to 0$, 
           the latter ones favouring a slightly larger $c_0\sim 0.20$.}
\la{1x1}
\end{figure}

Let us then discuss the approach to the continuum limit: $a\to 0$ 
and hence $\beta_G\to\infty$. 
The continuum extrapolation of $N_{1\times 1}$  is shown 
in Fig.~\ref{1x1}. It is seen that even though 
there is a lot of structure at a finite $\beta_G$, 
all the structure disappears when $\beta_G\to\infty$ and one 
gets a result which is independent of the parameters of the 
continuum theory~\cite{moorecomment}. 
In the continuum limit, the loop of size $1\times 1$ 
(as well as a loop of any finite size $n\times n$
in lattice units)
shrinks to a point, and it is clear that the result
is merely an artifact of the lattice regularization, and
sensitive only to UV-effects. 
Note that the continuum value $N_{1\times 1}\sim 0.2$
is close to the ``universal'' value seen, e.g., in~\cite{abh} 
(there it was obtained at a fixed lattice spacing but at the 
point of a second order phase transition, which corresponds 
precisely to the continuum limit). 

An important point to be noticed from Fig.~\ref{1x1} is that
in the broken phase ($x=0.0463,y=0.08$), the asymptotic 
$\beta_G$ regime is obtained quite late, $\beta_G\gsim 8$.
This is somewhat surprising since the smallest physical correlation 
length at this point is $1/m_W\sim 0.4/e_3^2$~\cite{prb}, corresponding to 
$\sim 2.4 a$ already at $\beta_G=6$. Thus one would expect to be
approaching the continuum limit earlier. The Higgs correlation
length is much larger, $1/m_H\sim 2/e_3^2\sim 12 a$ at $\beta_G=6$.

For larger loops $n\times n$, the qualitative
behaviour is similar to that for $N_{1\times 1}$, although the
numerical values are different. For large $\beta_G$,
including only terms linear in $a$, 
we expect the values of $N_{n\times n}$ to behave as  
\be
N_{n\times n} \approx f(n)+[d(x,y)+e(x,y) n]/\beta_G + {\cal O}(1/\beta_G^2). 
\la{expansion}
\ee
In principle, there could be a $\ln\beta_G$-term
in $d(x,y),e(x,y)$.
For fixed $n$, the continuum value is $f(n)$, but it
does not reflect the infrared
dynamics of the theory. Fitting the data, 
the functional form of $f(n)$ is found to be
consistent with $a+b/n+\bar c\ln n$  for large $n$. 
We cannot conclusively determine whether the coefficient $\bar c$ is 
non-vanishing or not. 
Assuming $\bar c = 0$, we get $a\sim 0.33, b\sim -0.13$, 
but if $\bar c$ is allowed to be non-zero, the absolute values of $a$ and $b$
are somewhat smaller. 
The numerical determination of $\bar c$ is difficult, since
it requires large values of $n$, for which very
large lattices are needed to remove the finite size effects. In any 
case, the real physics lies in the coefficient of 
$1/\beta_G$ (see below), 
for which a fit of the form in eq.~\nr{expansion} works 
very well (the confidence level is
CL=10--90\%, depending on the parameter values $x,y$).

\begin{figure}[t]
\hspace{1cm}
\epsfysize=8cm
\centerline{\epsffile{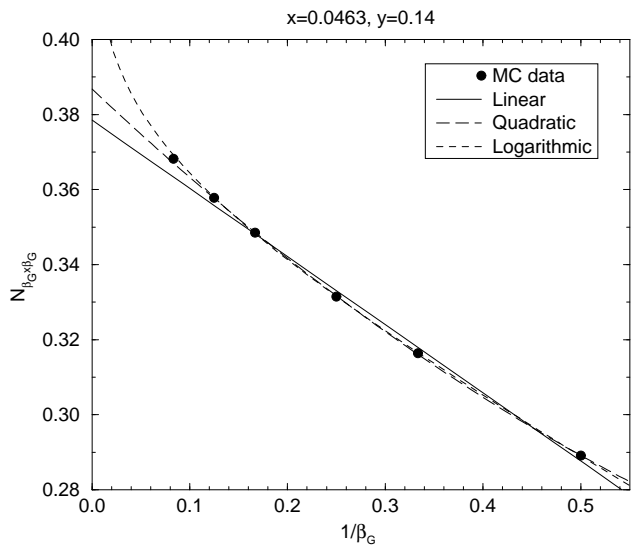}}
\caption[a]{The continuum extrapolation of $N_{\beta_G\times\beta_G}$
           at $x=0.0463,y=0.14$. Linear, quadratic and logarithmic
           fits are shown; the logarithmic one has the best confidence
           level, CL$=$11\%. The value of $N_{\beta_G\times\beta_G}$ 
           in the continuum limit is dominated by an unphysical 
           regularization sensitive constant part, see the text.}
\la{bxb}
\end{figure}

Let us then discuss loops which are of a fixed size 
in physical units: $( c/e_3^2)\times( c/e_3^2)$, 
where $c$ is a constant.
In lattice units this loop is $ c\beta_G\times  c\beta_G$, i.e., 
the size is varied as a function of $\beta_G$.
The continuum extrapolation for the loop $N_{\beta_G\times\beta_G}$
($ c=1$) is shown in Fig.~\ref{bxb}
at the point $x=0.0463,y=0.14$.
According to eq.~\nr{expansion}, we expect that
in the continuum limit,
\be
N_{( c\beta_G)\times( c\beta_G)}\approx \lim_{\beta_G\to \infty} 
f( c \beta_G)+  c e(x,y) + {\cal O}( c^2).
\la{physical}
\ee
The first term is unphysical and corresponds to the
regularization effects in eq.~\nr{condensate} below. (If there
is a term $\ln\beta_G$ in $e(x,y)$, 
then the regularization sensitive part
can also depend on $x,y$, but this dependence is analytic and
does not affect any phase transitions.) 
Thus the absolute value of $N_{( c\beta_G)\times( c\beta_G)}$ is 
not physical, only its changes are (see Fig.~\ref{dis}). 
The physical effects, i.e. $c e(x,y) + {\cal O}( c^2)$, 
come form the coefficient of $1/\beta_G$ 
(and ${\cal O}(1/\beta_G^2)$)
in eq.~\nr{expansion}.
To understand better the behaviour in eq.~\nr{physical}, 
let us discuss observables simpler than $n_C$.

Consider a typical
composite operator, such as $\langle\phi^*\phi\rangle$.
For $\langle\phi^*\phi\rangle$, one can make a perturbative computation
to find out what happens in the continuum limit.
It turns out the there is a linear (1-loop) 
and logarithmic (2-loop) divergence. The finite
$\msbar$-scheme continuum  result 
$\langle\phi^*\phi(e_3^2)\rangle_\rmi{cont}$
is \cite{lattcont}
\be
{\langle\phi^*\phi(e_3^2)\rangle_\rmi{cont}
\over e_3^2}=\fr12\beta_H\beta_G
\langle\phi^*\phi\rangle_\rmi{latt}-{3.1759115\beta_G\over4\pi}-
{1\over8\pi^2}\biggl[\log(6\beta_G)+0.668\biggr],
\la{condensate}
\ee
where ``latt" refers to the normalisation of the field in the
lattice action \nr{standardaction}. The second term 
on the RHS is the linear,
and the third the logarithmic divergence. The value of
$\langle\phi^*\phi\rangle_\rmi{latt}$ measured on the lattice
is thus not in itself a physical quantity: only its changes 
(such as the discontinuity across a first order transition) are,
since in them the divergent parts cancel. 

We now expect that a similar thing happens for $n_C$.
The difference is that $n_C$ is a more complicated 
and non-local quantity. The regularization sensitive part is 
not easily computable in perturbation theory, 
since even the integral 
\be
N_C^{\rm free} = 
\int {\cal D}\phi\,|n_C|
\exp\Bigl(-\int d^3x |\partial_i \phi|^2\Bigr)
\la{fst}
\ee
is not Gaussian, due to the non-polynomial
expression of $|n_C|$. A numerical evaluation
of $N_C^{\rm free}$ (a mass term has to be included
on the lattice to kill a zero mode) gives a result
close to the fitted values of
$f(n)$ in eq.~\nr{expansion} 
(and favours $\bar c=0$).

\begin{figure}[t]
\hspace{1cm}
\epsfysize=8cm
\centerline{\epsffile{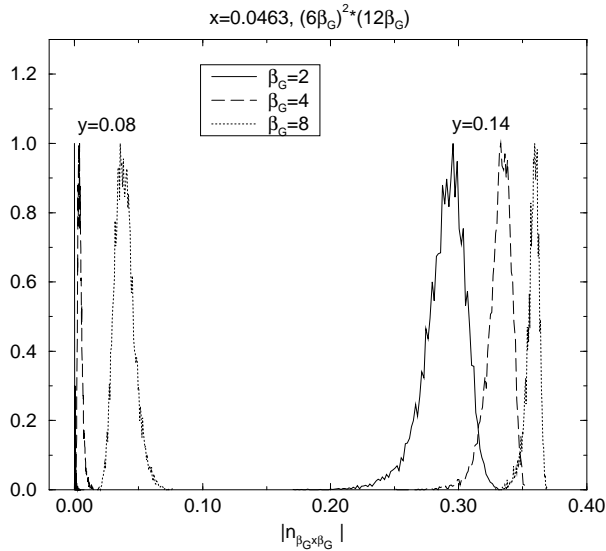}}
 
\caption[a]{The distributions of
the volume average of $|n_{\beta_G\times\beta_G}|$
for values of $y$ just below and above the critical point $y_c$
at $x=0.0463$. Precisely at $y_c$, the two peaks would appear
simultaneously.}
\la{dis}
\end{figure}

To demonstrate that the changes in $N_{\beta_G\times\beta_G}$
are physical, the distributions 
of $N_{\beta_G\times\beta_G}$  at $x=0.0463, \beta_G=2,4,8$ 
around the first order phase transition are shown in Fig.~\ref{dis}.
It is seen that the two-peak structure 
(in particular, the distance between the peaks)
indeed remains the same within statistical errors when $\beta_G$ is varied,
when $\beta_G$ is large enough ($\beta_G=4,8$). Thus the two-peak structure
is physical and has a finite continuum limit, while the location
of the structure on the $|n_{\beta_G\times\beta_G}|$-axis is 
unphysical and dominated by UV-effects: both peaks move to the right 
when $\beta_G$ increases (the location of the 
$y=0.14$ peak is shown in Fig.~\ref{bxb}). 
Note that $\beta_G=2$ is not yet in the scaling regime, and 
thus the distance between the peaks is different 
from that at $\beta_G=4,8$. 

For $x=2$, there is only a single peak which moves
continuously to larger values as $y$ is increased, 
see Fig.~\ref{yij}.

\section{Conclusions}

In conclusion, we have given a gauge-invariant definition for
a vortex passing through a loop on a lattice, measured the 
corresponding total vortex density, and discussed its extrapolation
to the continuum limit. We have pointed out that 
to approach a meaningful continuum limit, one must keep the 
size of the loop fixed in physical units. We have found that the
total vortex density behaves as a pseudo order parameter, 
analogously to $\langle\phi^*\phi\rangle$: 
the absolute value is always non-zero and is 
dominated by regularization effects near the continuum limit.
Thus only the changes of the total vortex density with respect
to the continuum parameters are physically meaningful. 
In the type I regime, the total vortex density displays
a discontinuity, whereas in the type II regime, it behaves
smoothly as the phase transition is crossed (Fig.~\ref{yij}).

The system possesses also (non-local) observables which
behave analogously to true order parameters. One such is 
the photon mass, which vanishes exactly in the symmetric
phase. It has been suggested that another such quantity
might be the density of long vortices passing through
the whole lattice~\cite{hulsebos,abh}. In contrast to the 
photon mass, this quantity should vanish in the broken phase
and remain non-zero in the symmetric phase.
The measurement of the density of long
vortices is in progress~\cite{a6}.

Finally, it would be interesting to study the spatial distribution 
of vortices. This can be done by measuring correlators of 
``vortex density operators'' $n_C$, separated by a distance $r$. 
One can define several correlators, depending on the 
relative orientations of the loops used in the $n_C$'s.
These quantities would give realistic initial conditions for 
simulations of the time evolution of vortex networks.

\section*{Acknowledgements}

A.R. would like to thank G.~Blatter and T.W.B.~Kibble
for discussions. We are grateful to G.D. Moore 
and K. Rummukainen for very
useful comments. The simulations were carried out with a 
Cray C94 at the Finnish Center for Scientific Computing.
This work was partly supported by the
TMR network {\em Finite Temperature Phase Transitions
in Particle Physics}, 
EU contract no.\ FMRX-CT97-0122.

\end{document}